\documentclass[final,1p,times]{elsarticle} 

\usepackage{graphicx}
\usepackage{amssymb} 
\usepackage{amsthm} 
\usepackage{lineno}

\newcommand\tauf{{\tau_{\rm f}}}
\newcommand\Tf{{T_{\rm f}}}
\newcommand\ds{{d_{\rm s}}}

\newcommand\pfrac[2]{\left(\frac{#1}{#2}\right)}
\newcommand\la{\left\langle\,}
\newcommand\ra{\,\right\rangle}

\newcommand\vs{{v_{\rm s}}}
\journal{Nuclear Physics A} 

\begin{document}

\begin{frontmatter} 


\title{Hydrodynamic noise and Bjorken expansion}

\author[a1]{J.~I.~Kapusta}
\author[a2]{B.~M\"uller}
\author[a3]{M.~Stephanov\fnref{presenter}}
\fntext[presenter] {Presenter}
\address[a1]{School of Physics \& Astronomy, University of Minnesota, Minneapolis, MN 55455}
\address[a2]{Department of Physics, Duke University, Durham, NC 27708-0305}
\address[a3]{Department of Physics, University of Illinois, Chicago, IL 60607}

\begin{abstract}
  Using the Bjorken expansion model we study the effect of intrinsic
  hydrodynamic noise on the correlations observed in heavy-ion
  collisions.
\end{abstract}

\end{frontmatter} 


Rich data on fluctuations and correlations in heavy-ion
collisions~\cite{Agakishiev:2011fs,Aamodt:2011by,Chatrchyan:2012wg,ATLAS:2012at}
raise many interesting questions, the answers to which could shed
light on the properties of the medium created in these collisions.
For example, it has been understood recently that long-range
rapidity-independent correlations are caused by the fluctuations of
the initial  collision geometry  which are propagated
hydrodynamically~\cite{Alver:2010gr}. We wish to
highlight the fact that in addition to the initial state fluctuations,
which are outside of the scope of hydrodynamics, there is an intrinsic
source of fluctuations in hydrodynamics itself --- local thermal
noise. We explore the effect of this noise on observed correlations in
a longitudinally expanding fireball created in heavy-ion
collisions~\cite{Kapusta:2011gt}.

What is the origin of the intrinsic hydrodynamic noise? The local
thermal equilibrium that underlies the hydrodynamic description is a
statistical concept. The thermodynamic state is not a single state,
but an ensemble of states, macroscopically similar, but with
microscopic differences fluctuating from one member of ensemble to
another member. For example, the equation of state expressing pressure
as a function of energy density, $P=P(\varepsilon)$, gives the {\em
  average} value of the pressure. The pressure fluctuates even if
the energy density is fixed (picture a gas of molecules in a box hitting
its wall). Had there been no fluctuations, correlators such as,
e.g., $\langle\,T_{\mu\nu}T_{\alpha\beta}\,\rangle$ (which give
viscosity via Kubo formula), would have vanished.

The relationship between fluctuations and dissipation suggests the
idea~\cite{Kapusta:2011gt} that measurements of fluctuations may yield
information about dissipative coefficients, such as viscosity,
complementary to existing methods based on the azimuthal asymmetry of
flow.

To see how the noise enters in hydrodynamic equations it is helpful to
consider the origins of hydrodynamics. We know that a perturbed fluid
relaxes back to an equilibrium. There are two very well
(parametrically) separated scales associated with this process. First,
the {\em local} thermal equilibrium is established on a relatively
short time scale characteristic of microscopic processes, such as
collisions. Achieving the {\em global} equilibrium, i.e., the same
thermodynamic conditions throughout the volume of the fluid, takes
considerably longer time, which grows as the square of the typical
size of non-homogeneities.  Hydrodynamics describes this much slower
process.

Hydrodynamics is an effective theory which deals only
with degrees of freedom that matter -- macroscopic quantities
characterizing the local thermal equilibrium such as energy, momentum
and charge densities, which change slowly because the
corresponding quantities are conserved. Faster microscopic degrees
of freedom left out of the hydrodynamic description are the noise.

To formalize the above picture we follow the approach of
\cite{statphys2} and apply it to a relativistically covariant
formulation of hydrodynamics.  In this formulation, the covariant
degrees of freedom are the components of the 4-velocity $u^\mu$ of the
local rest frame (the frame where the momentum density $T^{0i}$
vanishes) and the energy density $\varepsilon$ in this frame. We can
express 4 independent components of $T^{\mu\nu}$ in terms of the
variables $\varepsilon$ and $u^\mu$.

To close the system of equations the remaining 6 components of the
$T^{\mu\nu}$ (the stress in the local rest frame) must be also
expressed in terms of $\varepsilon$ and $u^\mu$. In equilibrium
$T^{\mu\nu}$ is constant throughout the system and Lorentz invariance
together with definitions of $\varepsilon$ and $u^\mu$ mean that the
stress tensor must be given by
\begin{equation}
T^{\mu\nu}_{\rm
  eq} = \varepsilon u^\mu u^\nu + P(\varepsilon) \Delta^{\mu\nu},
\label{eq:Teq}
\end{equation}
where $\Delta^\mu_\nu=\delta^\mu_\nu + u^\mu u_\nu$ is the projector
on the spatial hyperplane in the local rest frame ($\Delta^\mu_\nu u^\nu =
0$). The deviations from the equilibrium are due to gradients and to
the first order there
are two covariant expressions, corresponding to shear and bulk
viscosities:
\begin{equation}
 \Delta T^{\mu\nu} = -\eta\Delta^\mu_\lambda\left[\nabla^\lambda u^\nu+\nabla^\nu u^\lambda
-\frac23 g^{\lambda\nu}(\nabla\cdot u)\right] -  \zeta \Delta^{\mu\nu}(\nabla\cdot u).
\end{equation}

The main point is that the expression $T^{\mu\nu} =
T^{\mu\nu}_{\rm eq} + \Delta T^{\mu\nu}$ holds only on {\em
  average}. Both sides of this relation fluctuate. For every
member of the ensemble there is a discrepancy $S^{\mu\nu}$:
\begin{equation}
   T^{\mu\nu} =  T^{\mu\nu}_{\rm eq}+\Delta T^{\mu\nu}_{\rm visc} +
   { S^{\mu\nu}}.
\end{equation}
The discrepancy is the noise from the fast microscopic modes
and, therefore, the correlation functions of this noise are local on the
hydrodynamic scale:
\begin{equation}
\la { S^{\mu\nu}}(x){ S^{\alpha\beta}}(y)\ra\sim\delta^4(x-y).
\end{equation}

The magnitude is determined by the condition that the equilibrium
probability distribution of $\varepsilon$ is given by the number of
microscopic states, i.e., exponential of the entropy at that
$\varepsilon$. 
In particular, dissipation due to shear and bulk viscosities
must be matched by noise
\begin{equation}
  \la S^{\mu\nu}(x)S^{\alpha\beta}(y)\ra = 
2T\left[\eta\left(\Delta^{\mu\alpha}\Delta^{\nu\beta}+\Delta^{\mu\beta}\Delta^{\nu\alpha}\right)
+ \left(\zeta-\frac23\eta\right)\Delta^{\mu\nu}\Delta^{\alpha\beta}\right]
\delta^4(x-y).
\end{equation}

Although the noise is local, the correlations induced by it are propagated by
hydrodynamic modes over macroscopic distances.
Stochastic equations
\begin{equation}
 \nabla_\nu\left({ T^{\mu\nu}_{\rm eq}+\Delta T^{\mu\nu}_{\rm
  visc}}+{ S^{\mu\nu}}\right)=0
\end{equation}
solved for fluctuations of $\varepsilon$ and $u^\mu$ around a {\em static\/}
equilibrium state give well-known equilibrium correlation functions
(e.g., Kubo relation for viscosity).
Our goal is to apply these equations to determine
correlations in a non-equilibrium, expanding fireball.

The simplicity and symmetry of the Bjorken solution allows analytical
treatment.  Furthermore, in this work we only consider rapidity
dependence (i.e., we integrate/average fluctuations over transverse directions).

We write the hydrodynamic equations in Bjorken coordinates and
linearize in the fluctuations of $\varepsilon$ and $u^\mu$. The
typical structure of a correlation function at Bjorken
freezeout time, $\tauf$, is given by the integral over Bjorken rapidity $\xi$ and
the Bjorken time $\tau$ reflecting the fact that the noise, which is the
source of the correlations, exists at all space-time points:
\begin{equation}\label{eq:GG}
  \la \rho(\xi_1,\tauf) \, \rho(\xi_2,\tauf) \ra
= \frac{2}{A} \int\limits_{\tau_0}^\tauf
\frac{d\tau}{\tau^3} \frac{{\nu}}{\epsilon+P}
\left[
\int_{-\infty}^\infty\! d\xi\,{G}_{\rho}(\xi_1-\xi;\tauf, \tau)
{G}_{\rho}(\xi_2-\xi;\tauf, \tau)
\right]
 \, ,
\end{equation}
where we use convenient notation $\rho\equiv\delta s/s$ for the
relative entropy density fluctuation and $\nu\equiv (4\eta/3+\zeta)/{s}$ for
``longitudinal'' viscosity to entropy ratio.  The fluctuation $\rho$
at point $\xi_1$ is sourced by the noise fluctuation at each
point $\xi$ at earlier time $\tau$ propagated hydrodynamically via the linear
response Green's function ${G}_{\rho}(\xi_1-\xi;\tauf, \tau)$. The
same fluctuation is also propagated to point $\xi_2$ via
${G}_{\rho}(\xi_2-\xi;\tauf, \tau)$ and the correlation is the
averaged product
of the two fluctuations.

\begin{figure}[htbp]
\begin{center}
\includegraphics[width=0.29\textwidth]{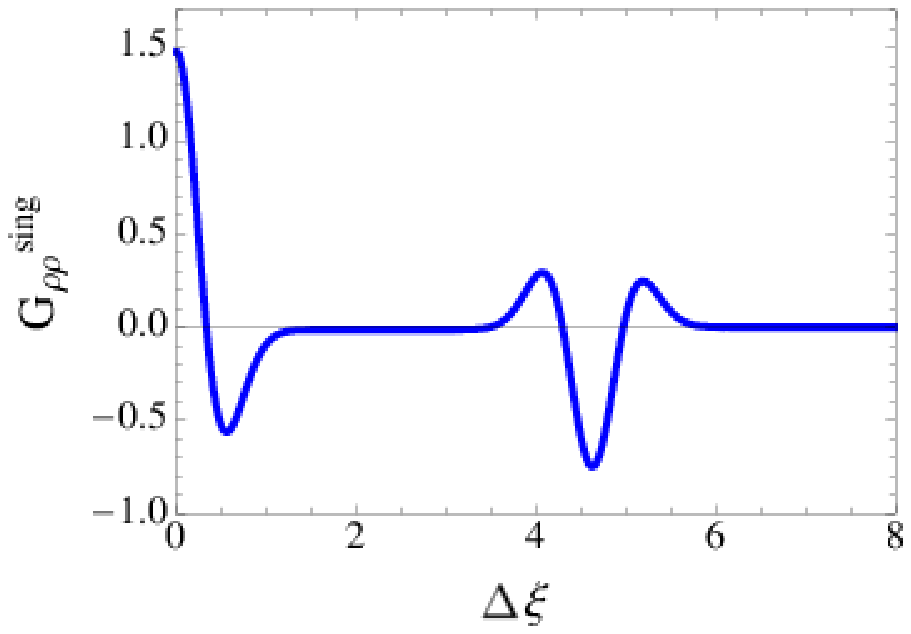}\qquad
\includegraphics[width=0.29\textwidth]{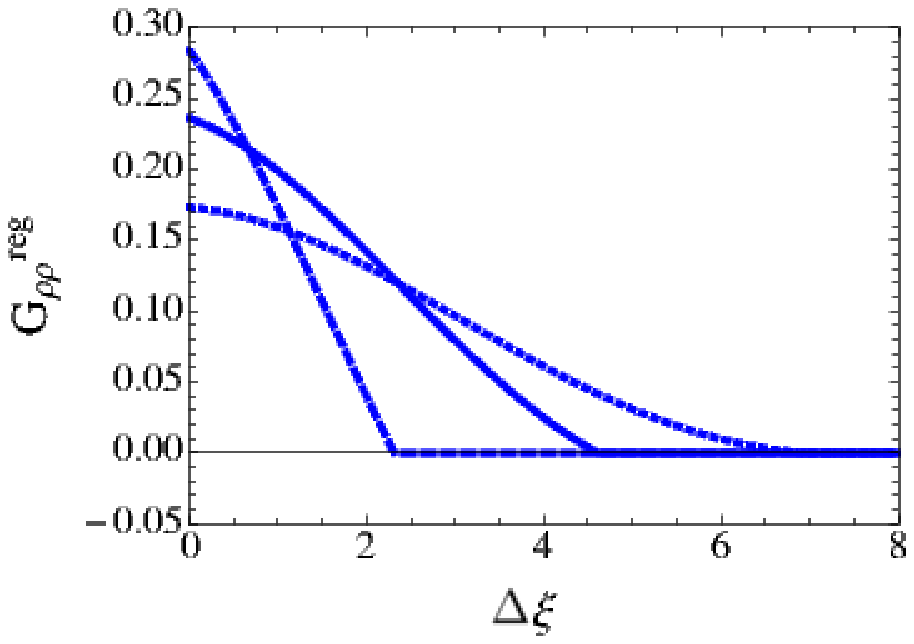}\qquad
\includegraphics[width=0.3\textwidth]{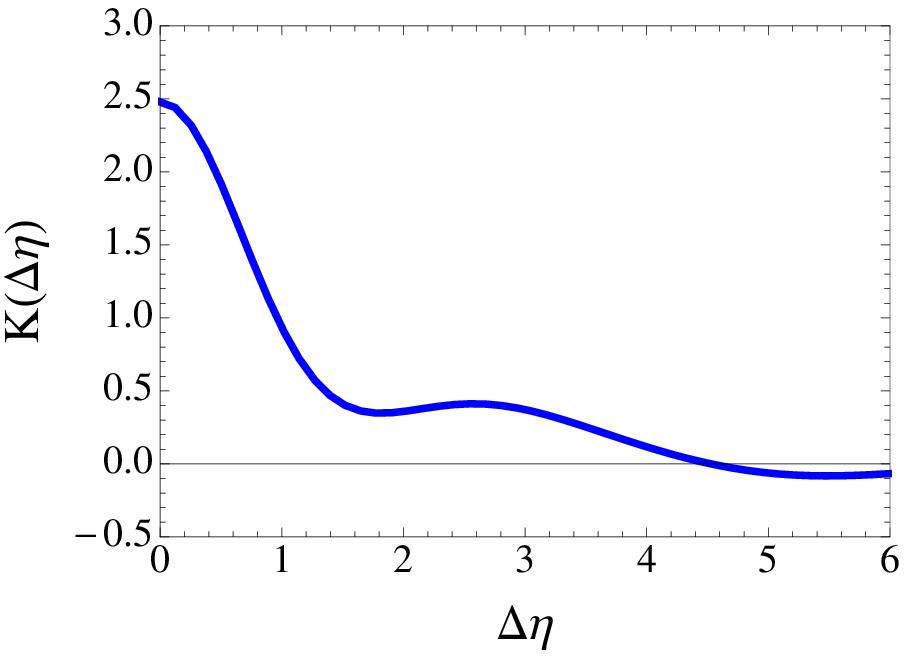}
\end{center}
\caption{{\it Left:} Singular part of the contribution of a single
  time slice $\tau$. ${\vs}^2=1/3$ and $\ln(\tauf/\tau)=4$. In order
  to visualize the infinitesimally narrow singularities we ``smear''
  the function with a Gaussian of square width $\sigma^2=0.1$. {\it
    Center:} The regular part of the correlation from 3 time slices
  $\tau$: $\ln(\tauf/\tau)=2,4,6$.
{\it Right:} Rapidity gap dependence of the
  particle-particle fluctuations in Eq.~(\ref{eq:dNdN/N}).}
\label{fig:Gsingreg}
\end{figure}

In order to demonstrate explicitly the contribution of each Bjorken
time slice $\tau$ 
to the final correlation function we consider separately the
expression in square brackets in Eq.~(\ref{eq:GG}), which we denote as
$G_{\rho\rho}(\Delta\xi;\tauf, \tau)$.  If one neglects viscosity, the
correlator $G_{\rho\rho}(\Delta\xi;\tauf, \tau)$ will have
singularities at the separation $\Delta \xi$ equal to twice the
distance traveled by sound on top of the expanding medium, namely,
$2\vs\ln(\tauf/\tau)$. This singular contribution is easy to
understand since it results from a noise fluctuation located exactly
halfway between the points $\xi_1$ and $\xi_2$ where correlation is
measured. We separate the singular contribution (defined as linear
combination of step function and its derivatives) and the remaining
regular (more precisely, continuous) contribution ${
  G}_{\rho\rho}\equiv G_{\rho\rho}^{\rm sing} + G_{\rho\rho}^{\rm
  reg}$ and plot them using characteristic values of parameters in
Fig.~\ref{fig:Gsingreg}{\it (left,center)}.

If the dispersion $\omega(k)$ of small fluctuations around the Bjorken
solution were linear, there would only be the sound-front singularities
at $\Delta\xi=2\vs\ln(\tauf/\tau)$ and 0. However,
$\omega=i\alpha\pm\sqrt{\vs^2k^2-\alpha^2}$ { [$\alpha\equiv
  (1-\vs^2)/2$]}. Note that the slowest mode behaves as $\omega_-\sim
-ik^2$ for $k\to0$. Correspondingly, the ``wake'' in
Fig.~\ref{fig:Gsingreg}{\it (center)} approaches a Gaussian for large
time intervals $\ln(\tau_f/\tau)$.  Although diffusion-like in
appearance, this mode is present even without dissipation.

The main effect of viscosity is, as usual, to smear the sound-front
singularity with a Gaussian of width which grows with time interval
between $\tauf$ and $\tau$  (initially following diffusive law $\sigma\sim\sqrt{\tauf-\tau}$):
\begin{equation}
\sigma^2 = \frac\nu\alpha\left( \frac{1}{\tau T} -
  \frac{1}{\tauf T_{\rm f}}\right)\,.\label{eq:1}
\end{equation}

Finally, we translate fluctuations of
hydrodynamic variables to observable fluctuations of particle
distributions $dN/d\eta$ using the Cooper-Frye freezeout formalism. It takes into
account the fact that particles within a given local subvolume at
Bjorken rapidity $\xi$ have kinematic rapidities $\eta$ which are
thermally spread around $\xi$. The result can be presented in the
following form:

\begin{equation}
  \label{eq:dNdN/N}
  \la \delta\frac{d N}{d\eta_1}\,\delta\frac{d N}{d\eta_2}\ra\la
  \frac{dN}{d\eta}\ra^{-1} = 
\frac{45\ds}{4\pi^4 N_{\rm eff}(T_0)}\,\frac{ { \nu}}{ \Tf\tauf}
\pfrac{{T_0^2}}{\Tf^2}^{\vs^{\!-2}-2} K(\Delta\eta) \,,
\end{equation}
where we separated the dependence on rapidity gap
$\Delta\eta=\eta_1-\eta_2$ into a factor $K(\Delta\eta)$. This factor
does not depend on other parameters such as  initial time and temperature $\tau_0$, $T_0$, the effective number of degrees of freedom
defined via $s(T)= 2 \pi^2N_{\rm eff}(T) T^3 /45$ and the mass
degeneracy of the observed species $\ds$ (e.g., 2 for charged pions).
It does depend on the ratio of the particle mass and the freezeout
temperature and for pions at $\Tf=150$ MeV it is plotted in
Fig.~\ref{fig:Gsingreg}{\it (right)}.

We conclude by describing a qualitative picture of the two-particle
correlations as a function of rapidity separation which may be helpful
in disentangling the two major contributions to correlations: initial
state fluctuations and hydrodynamic noise. The initial state
fluctuations, being fluctuations of the two-dimensional density of the
colliding Lorentz-contracted nuclear ``pancakes'', create correlations
which are largely independent of $\Delta\eta$ and stretch over the
whole rapidity interval between spectator fragments. In contrast,
hydrodynamic fluctuations are local, and it takes
time for them to spread over larger $\Delta\eta$. The effect of the
hydrodynamic noise is strongest at small $\Delta\eta\lesssim1$ leading to a
characteristic thermal peak (see Fig.~\ref{fig:Gsingreg}{\it
  (right)}). At larger $\Delta\eta$ the hydrodynamic fluctuations
induce correlations whose strength decreases with $\Delta\eta$ and
vanishes beyond the sound horizon at $2\vs\ln(\tauf/\tau_0)$. At such
large $\Delta\eta$ the correlation should reach a plateau set by
initial state fluctuations.

The magnitude  of the noise-induced
long-range correlations is proportional to
viscosity, Eq.~(\ref{eq:dNdN/N}). This fact could potentially be used
to measure or constrain the value of this important transport
coefficient by studying the  $\Delta\eta$ dependence of the
correlations. 

One can extend this analysis to the dependence of the correlations
on both rapidity and azimuthal angle separation ($\Delta\eta$ and
$\Delta\phi$) in a way similar to the $\Delta\phi$-only dependence analysis
in Ref.\cite{Staig:2011wj}. This work is in progress (see
T.~Springer's contribution to these proceedings,
\cite{Springer:2012iz}). Experimental data on $\Delta\eta$,
$\Delta\phi$ dependence of fluctuations, or, upon Fourier transform
with respect to $\Delta\phi$, the $\Delta\eta$ dependence of
$v_{nn}(\Delta\eta)$ could be compared to such a theoretical
analysis. Of course, a more realistic (e.g., 3d event-by-event
hydro~\cite{Schenke:2010rr}) simulation may be necessary to enable a
quantitative comparison and to determine or constrain the value of
the viscosity.

This work was supported by the U.S. DOE Grants No. DE-FG02-87ER40328 (JIK), 
DE-FG02-05ER41367 (BM), and DE-FG02-01ER41195 (MS).



\end{document}